# The Security of the United Kingdom's Electricity Imports under Conditions of High European Demand

Anthony D Stephens and David R Walwyn
Correspondence to tonystephensgigg@gmail.com

**Abstract**

Energy policy in Europe has been driven by the three goals of security of supply, economic competitiveness and environmental sustainability, referred to as the energy 'trilemma'. Although there are clear conflicts within the trilemma, member countries have acted to facilitate a fully integrated European electricity market. Interconnection and cross-border electricity trade has been a fundamental part of such market liberalisation. However, it has been suggested that consumers are exposed to a higher price volatility as a consequence of interconnection. Furthermore, during times of energy shortages and high demand, issues of national sovereignty take precedence over cooperation. In this article, the unique and somewhat peculiar conditions of early 2017 within France, Germany and the United Kingdom have been studied to understand how the existing integration arrangements address the energy trilemma. It is concluded that the dominant interests are economic and national security; issues of environmental sustainability are neglected or overridden. Although the optimisation of European electricity generation to achieve a lower overall carbon emission is possible, such a goal is far from being realised. Furthermore, it is apparent that the United Kingdom, and other countries, cannot rely upon imports from other countries during periods of high demand and/or limited supply.

**Key Words**

1. **Introduction**

Germany, France and the United Kingdom (UK) have the largest electricity generating capacities in Europe, but only Germany and France generate significant export surpluses [1]. Although for many years the UK also had substantial reserves of generating capacity, these have been depleted in recent years by the closure of many coal-fired power stations without establishing alternative capacity [2]. In October 2016 the National Grid predicted that the gap between the UK's total generating capacity and its peak winter demand could fall to only 1.1% [3].

Since 2007, European energy systems, including the UK, have been shaped by an energy policy directed at the common goals of security of supply, economic competitiveness and environmental sustainability, collectively referred to as the energy 'trilemma' [4, 5]. Although there are clear policy conflicts within the trilemma, and particularly between energy security and competitiveness [6], where the latter is closely linked to market liberalisation, countries within the European Union have acted to facilitate a fully integrated European electricity market [7]. Interconnection between European countries and cross-border electricity trade has been a fundamental part of such market liberalisation [8]. For instance, in 2014 the European Commission requested that member states should install an interconnection capacity of at least 10% of their total installed national capacity by 2020 [9].

A 2 gigawatt (GW) interconnector between the UK and France and 1 GW interconnector between the UK and Holland have in recent years run at close to full capacity, importing electricity to the UK [10], despite inefficiencies in respect of both prices and capacity utilisation as a consequence of abuses by the dominant generators [11]. Commentators on energy matters have expressed concerns that reliance on the European energy system might reduce the security of the UK's system during periods of high demand [12]. However, an analysis of potential vulnerabilities is generally difficult since severe stresses on the energy system, which are required in order to investigate its weaknesses, happen only infrequently. Furthermore, until relatively recently, real time data was not available to support the analysis of such events as they took place.

In January 2017, a number of factors which we shall discuss in the next section, combined to provide a useful stress test for the European energy system. This report will use the real-time records of Germany [13], France [14] and the UK [10] to analyse the performance of the European energy system during this period. It will be found that the records provide useful insights into the operation of the UK's interconnectors at a time of high stress on the European energy system, reflecting on how European energy policy works in practice rather on paper.

2. **Stresses on the European Energy System during the Winter of 2016/17**

The stresses which existed simultaneously within European generating system during the period in questions, disrupted normal energy flows. Further details on each stress now follow.

**2.1  Low Wind Across Western Europe**

Figure 1 shows a weather synoptic on 23$^{rd}$ January which was typical of the period between 15$^{th}$ and 25$^{th}$ January 2017, when there was a stationary high pressure area over most of Western Europe. Figure 2 shows wind energy generated in the UK, France and Germany during January 2017, when the wind patterns were largely in phase with one another.

**Figure 1: European weather synoptic for 23 January 2017**

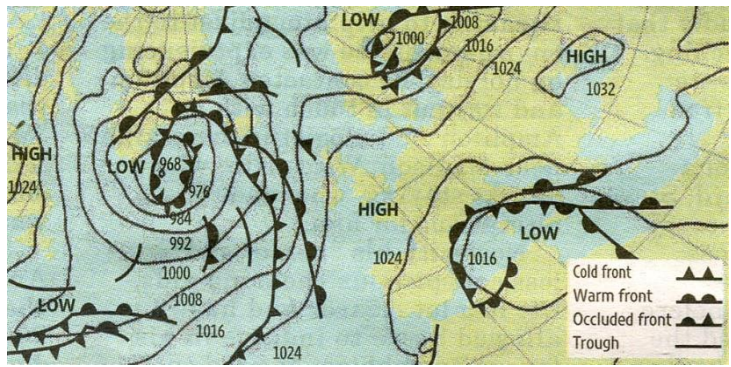

**Figure 2: UK, France and Germany wind generation in January 2017**

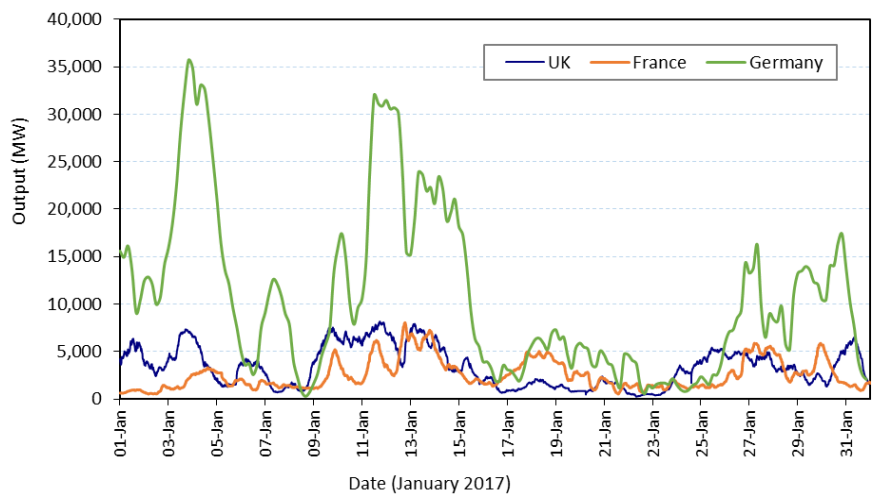

Source: Gridwatch [10], Fraunhofer Institute for Solar Energy Systems ISE [13], Gridwatch [14]

**2.2    High Energy Demand within Europe (including the United Kingdom)**

Unlike the UK, which uses mainly gas for domestic heating, France relies predominantly on electricity. As may be see in Figure 3, an exceptional cold spell across Europe resulted in electricity demand in France in January 2017 being approximately 10 GW above its 2016 level.

**2.3    Restrictions on Nuclear Generation in France**

Th main contributor to French electricity supply is its 58 nuclear reactors. It was therefore a serious matter when the French Nuclear Safety Authority (ASN), launched an investigation in June 2016 into concerns that French nuclear pressure vessels might have been illegally certified, raising serious doubts about the integrity of the in-service reactor fleet [15]. ASN required 20 of France's 58 nuclear reactors to be shut down for inspection [16] and in October 2016 France's exports of electricity had fallen to their lowest level since a period of cold weather in February 2012, the last occasion on which France had imported electricity from the UK [17]. This had caused electricity prices in the UK to rise above Euro 100 per

MWhr on a number of occasions during October. Figure 4 reveals a significantly lower level of nuclear generation for much of January 2017 compared with January 2016.

**Figure 3: Electricity demand in France in Jan 2016 (left) and Jan 2017 (right)**

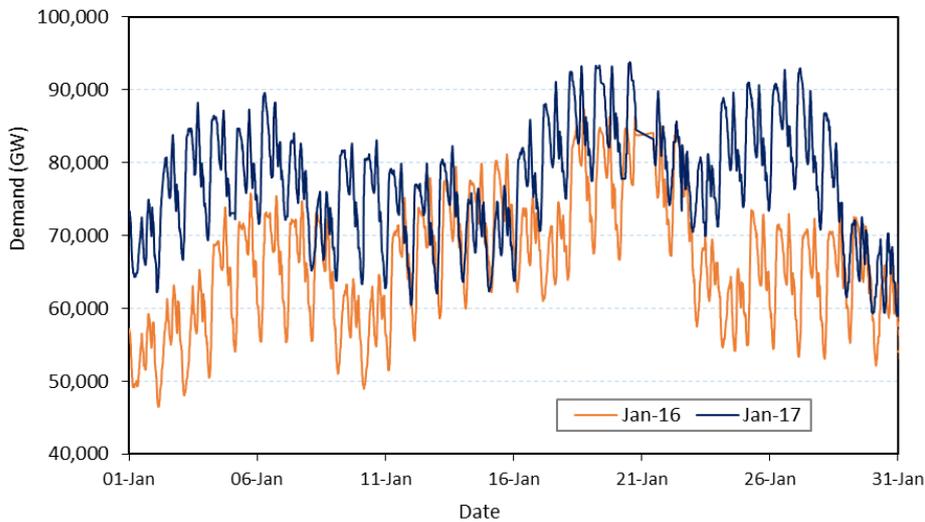

Source: Gridwatch [10]

**Figure 4: Nuclear generation in France in Jan 2016 (left) and Jan 2017 (right)**

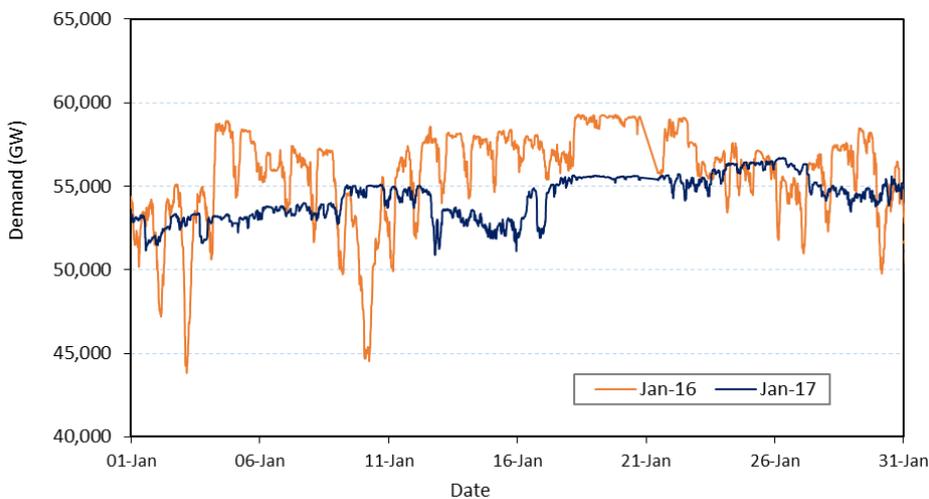

Source: Gridwatch [14]

### 2.4　Damage to the United Kingdom-France Interconnector

It was reported on 30[th] November 2016 that 4 of the 8 cables running along the sea bed between France and the UK had been damaged, possibly by the dragging of a boat's anchor, and that the damage cables would not be mended until February 2017 [3]. The article suggested that the tightening of supplies was likely to lead to increased volatility and higher energy prices in January and February, with the possibility of mothballed coal fired stations having to be pressed into service to meet electricity demand. The consequence of this damage reduced the interconnector's capacity from ± 2 GW to ± 1 GW, as shown in Figure 5.

**Figure 5: Usage of the UK–France interconnector in January 2016, and January 2017**

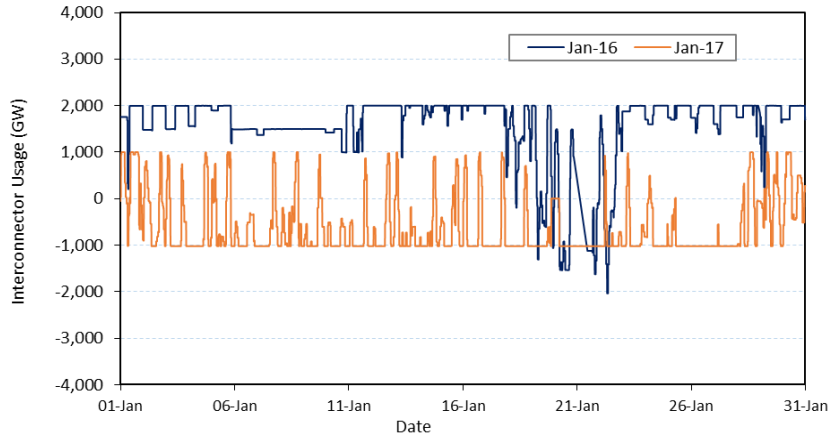

Source: Gridwatch [14]

### 3. Effect of the Stresses on United Kingdom Imports

The combination of raised domestic demand and reduced nuclear generating capacity resulted in a severe reduction in electricity available for export from France. As may be seen in Figure 6, France exported around 10 GW for most of January 2016, but was a net importer for much of January 2017. For most of January 2017, France imported 1 GW from the UK, in contrast with January 2016 when France mainly exported 2 GW to the UK.

**Figure 6: French electricity exports in 2016 and 2017 (negative values are exports and positive values imports)**

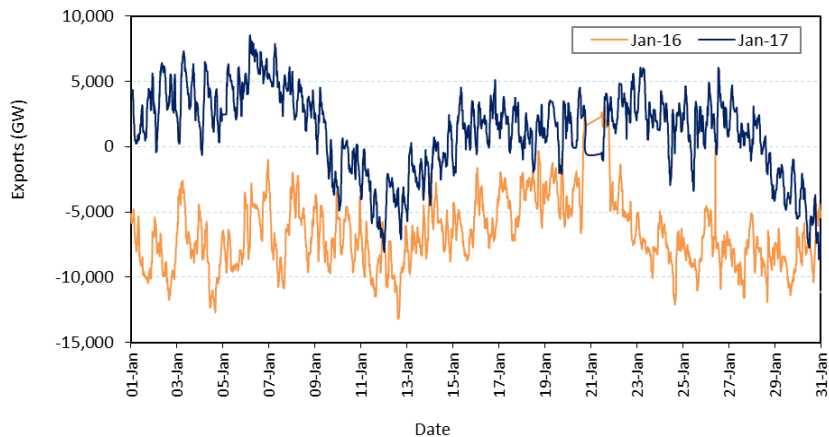

Source: Gridwatch [14]

The main source of electricity imported into the UK from Holland via the UK-Holland interconnector is Germany. We shall see later that not only was the German wind fleet unreliable during much of January 2017, but it also received less than normal imports from France. As may be seen in Figure 7, France was a consistent supplier of electricity to Germany in January 2016 but France mainly imported electricity from Germany over the same period in 2017.

**Figure 7: Exports from France to Germany in January 2016 and January 2017 (negative values are exports and positive values imports)**

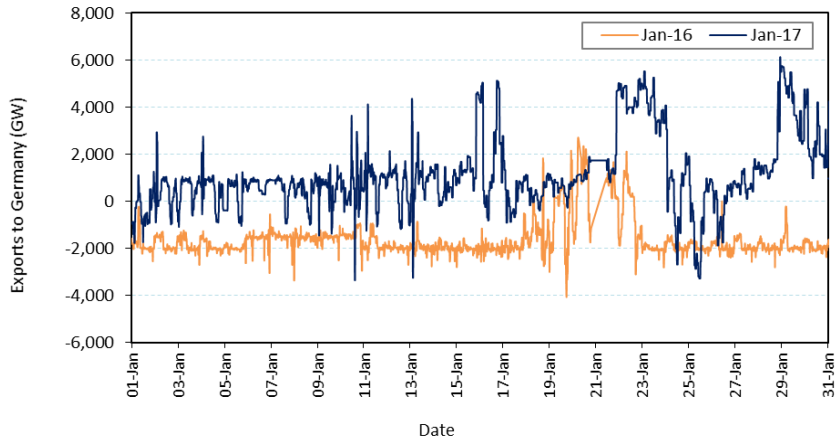

Source: Fraunhofer Institute for Solar Energy Systems ISE [13]

Although Germany's wind fleet, which has a capacity of 45 GW, generated 35 GW on 3$^{rd}$ January, its output during the month was highly variable. On 8$^{th}$ January, the wind fleet was only able to generate a total 0.3 GW, and during the 10-day period of high pressure throughout Europe (15$^{th}$ to 25$^{th}$ January), Germany's wind generation was negligible (see the green component in Figure 8). The highlights in this figure cover the weekends, when Germany's total demand was typically 20 GW below week-day demand, and also those periods during the month when wind generation in Germany was greater than 10 GW. The weekends and periods of high winds in Germany are also highlighted on the UK-Holland interconnector records of Figure 9. It can be seen that the UK-Holland interconnector was only a reliable source of power for the UK at weekends or during periods when German wind generation was greater than 10 GW.

**Figure 8: Main components of Germany's electricity generation in January 2017**

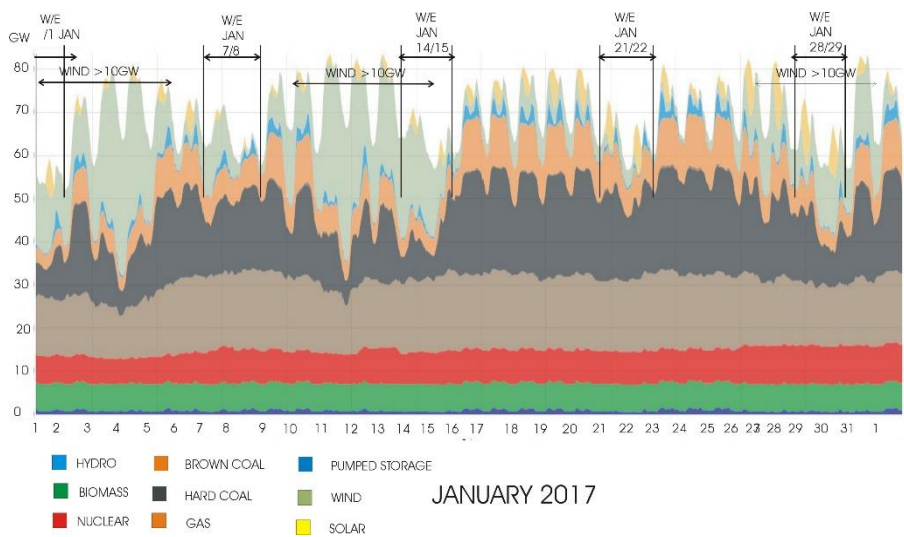

Source: Fraunhofer Institute for Solar Energy Systems ISE [13]

**Figure 9: Imports from Holland to the UK in January 2017 (highlighted are weekends and periods when wind generation in Germany was greater than 10 GW)**

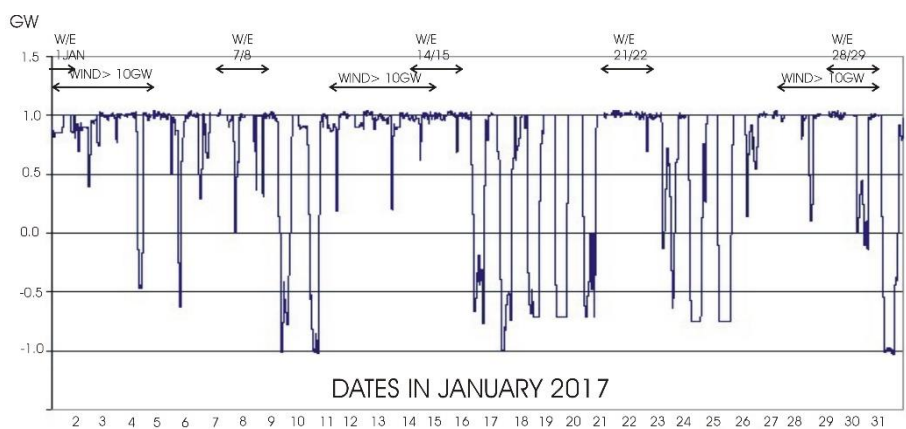

Source: Gridwatch [10]

This data on the usage of the UK's two electricity interconnectors with Europe in January 2017 confirms that it ought not to rely on the European energy system at times when the latter is under stress. France and Germany, the two countries in Western Europe which in normal times have significant energy export capability, will concentrate on satisfying their domestic demands at such times.

## 4. Discussion
### 4.1 Prospects for United Kingdom's Electricity Imports

Prior to the winter of January 2017, the previous occasion on which the UK was not able to rely on the UK-France interconnector for power in winter was in 2012. For reasons enumerated below, the European energy system is likely to have significantly less surplus generating capacity in future winters, making imports via the interconnectors less reliable.

1. Électricité de France (EDF) currently faces the serious problem that its 58 ageing nuclear reactors, which supply 75% of French electricity, are reaching the end of their lives. France's National Audit Office estimates that EDF will have to spend Euro 100 billion over the next 13 years to bring the reactor fleet up to modern safety standards [18]. If this is correct, France will inevitably have significantly less electricity to export than in the past. Indeed, during periods of high demand in winter France may become an importer rather than an exporter of power.

2. In addition to France's problems with its ageing reactor fleet is the problem of finding in 2015 elevated carbon content in the reactor pressure vessel of EDF's new flagship EPR reactor currently being built at Flamanville. It appears that the pressure vessel safety tests may have been deliberately faked by the Burgundy manufacturer over a period of several decades, calling into question the integrity of the entire reactor fleet. EDF was ordered to undertake emergency tests on 18 of its 58 reactors [16] and, despite the COP21 carbon emissions agreement which recently went into force, France had to revert to burning coal at its highest level in 32 years [19]. French exports of electricity in October 2016 were 89% below their October 2015 level, the most serious effect on prices being not in France but in the UK [17]. The seriousness of the problem resulted in RTE, the French grid operator, drawing up emergency plans to impose 2 hour power cuts on French homes and businesses [16].

3. The problems with the Flamanville EPR raise concerns about the viability of EDF's proposed £18billion EPR at Hinkley Point in Somerset, whose purpose is to provide 7% of Britain's electricity. Even if the Flamanville EPR comes on line in 2018, it will have taken 11 years to construct and, at Euro 10.5 billion, will be three times over budget [20]. With debts of Euro 37billion and renovation costs of Euro 100 billion for its reactor fleet many have doubted the ability of EDF to finance Hinkley Point. Those questioning the viability of the Hinkley Point EPR include EDF's former finance director who resigned arguing that EDF *could not afford to be pursuing expensive new overseas projects while financing such big challenges at home* [2].

4. Germany is becoming an increasingly unreliable supplier of energy for exports following its decision after the Fukushima nuclear disaster to phase out its own nuclear generation by 2022 [21]. Germany's nuclear generating capacity, which had been 20.43 GW in 2010 was reduced to 10.80 GW in 2016, and it will clearly have serious implications for the overall balance of energy in Western Europe should the remaining capacity be lost without replacement by 2022 [21].

The above threats to reliable supply from France and Germany are exacerbated by the decline in the UK's domestic generating capacity over recent years, a decline that is the result of the government's decision to phase out coal fired generation by 2025 without making provision for replacement with reliable sources of generation. Many organisations, including the Confederation of British Industry, Scottish Power and the Institution of Mechanical Engineers, have argued strenuously but unsuccessfully with government ministers for the building of new gas fired generation capacity [2, 22]. What the UK government put in place instead were capacity auctions held each December when generators were invited to bid for payments to provide standby winter capacity. The government's hope was that these payments would boost investment in new gas generating capacity. However, with energy prices at historic low levels, the auctions in 2015 and 2016 resulted in contracts being given mainly for the dirtiest fuels, namely diesel and coal fired generation [23, 24]. Keith Anderson, chief corporate officer of Scottish Power argued strongly against the concept of capacity auctions, stating that "*what the UK does not need is another 7 GW of diesel generation*" [2]. He also argued in favour of domestic generation rather than imports, stating "*as an economy we should be taking control of our own energy supply*", and supply should be from gas generation rather than new nuclear - "*as a country we have been saying we are committed to new nuclear for ten years. How long is (Hinkley Point) going to take to come through? Let's get on and build what we know how to build. The risk is we are sitting here in five years time and have not built anything*" [2].

What is astonishing about the UK government capacity auction strategy, which has resulted in the promotion of diesel and coal burning during winter months when atmospheric conditions can give rise to life threatening levels of pollution, is that it is completely at variance with the government's climate change strategy and EU energy policy. It is difficult to understand a strategy of investing heavily in green energy and at the same time becoming more reliant on diesel and coal fired generation. The Office of Budget Responsibility estimate that the cost of the government's green energy policies will have escalated to £11.4 billion per annum by 2020/21, a cost which will be borne by the UK's households at an average of roughly £438 per household per annum [25]. Combined Gas Turbine Generators on the other hand, which can be relied upon when there is no wind, are relatively quick to construct, and cost around £1 billion per installed GW. Gas generation would not only increase the UK's winter margins but also would reduce emissions to the environment, since gas generates only about half the carbon dioxide emission of coal.

Former Chancellor of the Exchequer, Lord Darling, speaking to a House of Lords Economics Affairs Committee, sharply criticised a plethora of government subsidy schemes, stating that "*Britain's energy market has become so heavily distorted by subsidy payments to suppliers of low carbon, renewable and nuclear power that it is now completely opaque…. The UK energy market is failing to deliver value for money to consumers*" [26]. Dermot Nolan, Chief Executive of Office of Gas and Electricity Markets, speaking to the same committee, blamed subsidy schemes designed to support the development of wind farms, solar panels and nuclear power stations including the £18 billion Hinkley Point plant. He was deeply sceptical of subsidies. A problem with the opaque economics of energy policy to which Lord Darling referred is that they suppress public discussion.

An article in the Economist in 2015 asserted that Germany was following the same combination of incoherent energy policies as the UK [27]. On the one hand Germany's subsidies for green energy were Euro 20 billion per annum, the highest in Europe, but it was actually increasing the burning of "dirty" hard and brown coal. Because of the lack of wind, pollution was frequently a serious health hazard throughout Europe during the early winter of 2016/16, with public recriminations between France and Germany about which country was to blame [28]. It is therefore surprising that Germany did not always take the opportunity to minimise coal burning when it had the opportunity to do so. On the week-end of 21/22 January, when German domestic demand fell below the level during the previous week, it chose to use spare coal generating capacity to export 1 GW of electricity to the UK via Holland, although it had restricted its exports to the UK during the week.

**Figure 10: Components of UK generation during January 2017**

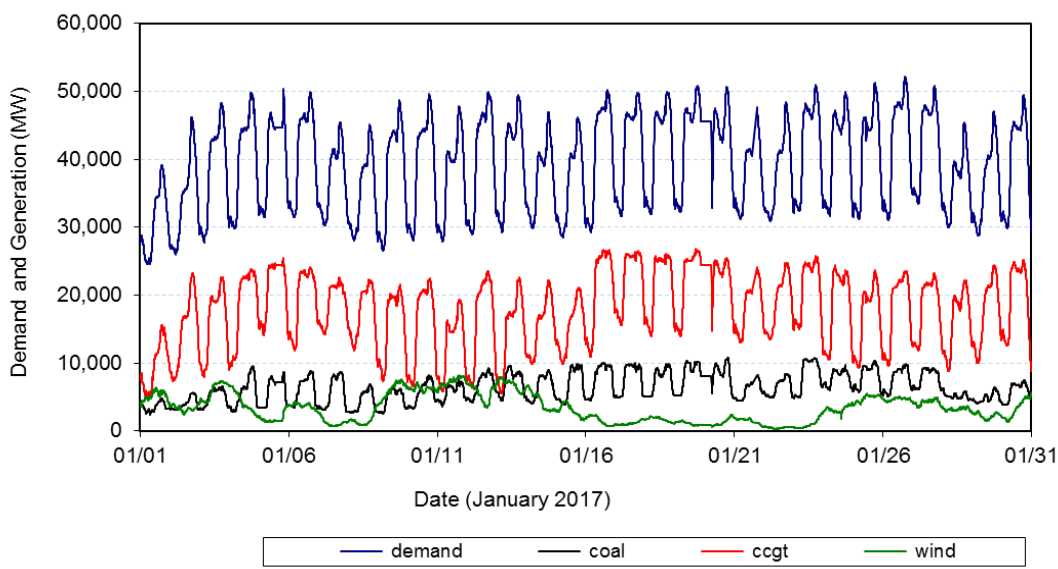

As may be seen in Figure 10, the UK did not need this 1 GW from Germany on the weekend of 21/22 January, since it had reduced its own generation below the level of the previous week. The UK had sufficient spare gas capacity to generate an additional 1 GW, obviating the need to import 1GW from Germany. In burning mainly coal to generate the 1 GW for export, Germany will have generated an additional 56,800 tonnes of carbon dioxide emissions that week-end, while 1 GW generated from gas in the UK would only have generated 23,400 tonnes of carbon dioxide emissions (a net reduction of carbon dioxide emissions to the environment of 33,400 tonnes). Indeed, if the objective were to minimise atmospheric pollution, rather than Germany export 1 GW to the UK, the UK should have exported 1 GW to Germany via Holland, enabling Germany to reduce is coal generation by another GW.

**Figure 11: German production records and exports in January (exports are shown below the line)**

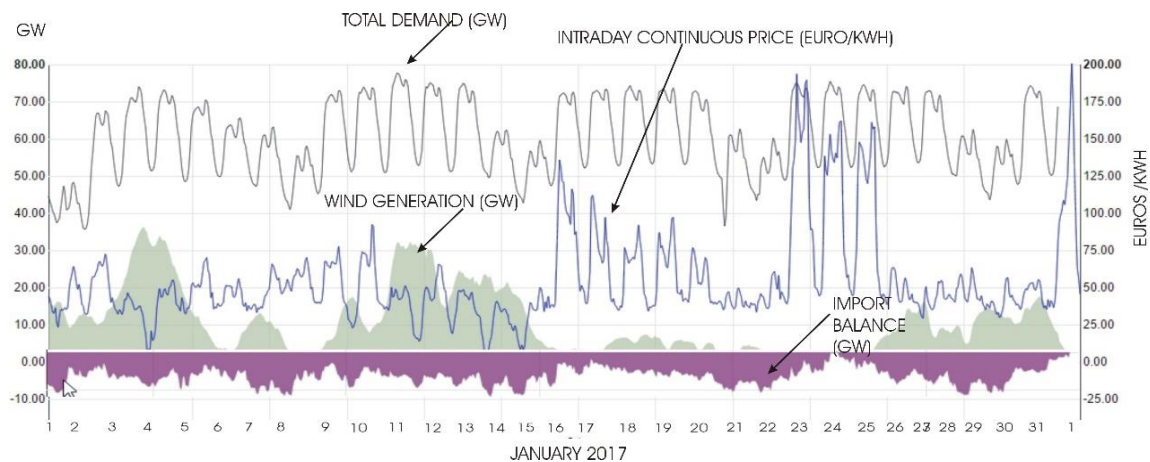

Source: Fraunhofer Institute for Solar Energy Systems ISE [13]

It is interesting to see how German demand and wind generation impacted so strongly on prices and exports in January 2017. A combination of high wind and low demand led to high exports but at prices close to zero. In contrast, high demand and low wind generation led to extremely high prices. On one occasion during the period when wind generation was low throughout Europe (15$^{th}$ to 25$^{th}$ January 2017), German electricity prices reached nearly Euro 200 per kWh, as shown in Figure 11. During periods of low wind German exports are generally below their normal level except at weekends, when prices fall because of reduced demand. Thus, during the week-end of 21$^{st}$/22$^{nd}$ January 2017, an average of 7 GW was exported from Germany (1 GW of which went to the UK), generated mainly from coal.

Clearly exports are currently driven by price, without any environmental considerations, price being determined by supply and demand. Particularly during periods of high pollution, export strategy ought to take into consideration environmental factors. A general principle might be that if two countries have spare generating capacity, any exports between them should be from the country with the lower emissions based on its surplus generation capacity. Such exports may be indirect through a third country, providing the end result is a net reduction in emissions. Thus, although France's nuclear generation creates significantly less emissions that gas generation, it would be valid to export gas generation from the UK through France for onwards transmission to Germany, (where it would displace coal generation). Such a rule would be easy to administer, since emission characteristics (e.g. kg carbon dioxide generated per kWhr) may be calculated in real time from data which is already available on line.

### 4.2    Future Use of Interconnectors between the UK and Europe

Perhaps stimulated by a lack of generating capacity in the UK and surplus generating capacity in France, plans are already in place to enhance regional cooperation and build several new interconnectors in the near future [7]. It is anticipated that by the early 2020s an additional 6.4 GW of interconnectors capacity will link the UK with France [29, 30]. Furthermore similar additions to existing interconnectors are planned within Europe [31], with studies showing that improvements to interconnectivity will be critical for competitiveness and the need to accommodate variable renewable energy generation [32].

An interesting question is how this additional interconnector capacity will be used, given that the capacity is unlikely to be useful to the UK during severe winter shortages, the problem being lack of generating capacity rather than lack of interconnectors. It is also possible that the current problems in France will result in France having significantly less surplus to export in future, questioning the economic viability of these additional links (capital costs for the links are typically around £1 billion per 2GW capacity).

It was at one time argued that investments in interconnectors would enable wind surpluses to be transmitted across Europe from countries in surplus to those in deficit. The problem with this argument however is that, as in January 2017, wind systems are reasonably well coordinated across the main generating and consuming countries in Europe. There is also a problem of how and where to send large surpluses when the wind is blowing strongly. Although there was practically no wind generation between $15^{th}$ and $25^{th}$ January, Germany briefly generated 30 GW on two occasions in the month, but there were also high winds in both the UK and France at the same time. When all the UK wind turbines currently under construction or having received consent are fully operational in the UK, the UK should have around 25-30 GW of wind generating capacity. When this point is reached, there are likely to be few occasions when it will be economic to export large wind surpluses between Germany and the UK. Firstly, the surpluses will largely occur at the same time in both countries and, secondly, it will never be economic to invest in interconnectors costing £1 billion per 2GW to handle large intermittent short term surpluses. The logic would appear to be for wind generation to be used mainly in its country of generation, and for interconnectors to transmit mainly base load generation between countries.

In 2016, Stephens and Walwyn [33] looked at the consequences of the UK limiting its wind generation to that which it could economically use. When the wind fleet is small all wind generation can be usefully used, but as it increases in size it reaches a point at which wind has to be spilt, causing the load factor of the wind fleet to fall. To quantify the relationship between the size of the wind fleet and its load factor, Stephens and Walwyn [33] modelled this correlation based on UK electricity generation records for the years 2013, 2014 and 2015. Separate models were built using the data for each of the three years. The models suggested that up to a wind fleet capacity of around 25 GW, all the wind generation is usefully used and the load factor of the wind fleet is around 32.5%. Above that size, wind has to be progressively shed and a 40 GW wind fleet is anticipated to have a marginal load factor of only around 18%. The conclusion reached was that 25 GW is approximately the upper economic limit for a UK wind fleet, when the wind fleet will be generating approximately 20% of the UK's demand. This provides us with an estimate of the economic upper limit of a UK wind fleet and also suggests an upper limit on the wind fleet's ability to decarbonise the grid without the use of storage technologies.

5. **Conclusion**

The UK, German and French electricity generation records reveal that a number of unusual factors combined in January 2017 to severely reduce generation capacity in Europe. Despite the fact that the UK-France and UK-Holland interconnectors have provided the UK with about 7% of its electrical power reliably for several years, the experience in January 2017 was that interconnectors cannot be relied upon when there is a general shortage of power in Europe. Germany and France will give their own domestic markets priority. For reasons outlined in the paper, shortages of generating capacity in Europe are likely to become much more frequent in future years and, as a matter of some urgency, the UK needs to put in place strategies to reinforce its generating capacity.

Although European energy policy claims to be based on the three equally-important goals of security of supply, economic competitiveness and environmental sustainability, it is apparent that in practice environmental sustainability, and to some extent economic competitiveness, remain of lower priority within the trilemma. Environmental concerns seem to be particularly neglected. With winter pollution now becoming a more serious problem in Europe, the role that interconnectors could play a role in minimising overall emissions to the environment is substantially underutilised. At times of high pollution risk, it is proposed that power should be raised in countries able to generate electricity with the least pollution (measured in kg carbon dioxide per kWhr generated), exporting to countries creating more emissions. In practice, this means favouring exports from nuclear and gas generation rather than coal generation.

The paper summarises the reasons for interconnectors not being an appropriate means of transmitting large wind surpluses around Europe. In the main, wind generation should be used in its country of origin. The results of modelling work on UK generation data suggests that an upper economic limit on a UK wind fleet is likely to be around 25 GW [33]. A project has recently started, using the same approach as in the UK, to investigate the likely upper limit of the German wind fleet.